\documentclass[11pt]{article}
\begin{document}
\title{Current Algebra Associated with Logarithmic Conformal Field Theories}
\author{S. Moghimi-Araghi \footnote{e-mail: samanimi@rose.ipm.ac.ir} ,
S. Rouhani \footnote{e-mail: rouhani@karun.ipm.ac.ir} and M.
Saadat \footnote{e-mail: saadat@mehr.sharif.ac.ir}\\
\\
Department of Physics, Sharif University of Technology,\\ Tehran,
P.O.Box: 11365-9161, Iran\\ Institute for studies in Theoretical
physics and Mathematics,\\ Tehran, P.O.Box: 19395-5531, Iran}
\maketitle
\begin{abstract}
We propose a general frame work for deriving the OPEs within a
logarithmic conformal field theory (LCFT). This naturally leads to
the emergence of a logarithmic partner of the energy momentum
tensor within an LCFT, and implies that the current algebra
associated with an LCFT is expanded. We derive this algebra for a
generic LCFT and discuss some of its implications. We observe
that two constants arise in the OPE of the energy-momentum tensor
with itself. One of these is the usual central charge.
\vspace{5mm}
\end{abstract}

Since the paper by Belavin, Polyakov and Zamolodchikov \cite{BPZ}
on the determining role of conformal invariance on the structure
of two dimensional quantum field theories, an enormous amount of
work has been done on the role of conformal field theories (CFTs)
in various aspects of physics such as string theory, critical
phenomena and condensed matter physics. A decade later, Gurarie
\cite{Gur} pointed to the existence of LCFTs. Correlation
functions in an LCFT may have logarithmic as well as power
dependence. Such logarithmic terms were ruled out earlier due to
requirements such as unitarity or non existence of null states.
The literature on LCFT is already very long, for a survey see
some of the recent papers on LCFT for example
\cite{FlohrNew,Lewis,Kogan}.

In an LCFT, degenerate operators exist which form a Jordan cell
under conformal transformations. In the simplest case one has a
pair $C$ and $D$ transforming as:

\begin{eqnarray}
C(\lambda z)&=&\lambda^{-\Delta}C(z) ,\nonumber\\
D(\lambda z)&=&\lambda^{-\Delta}[D(z)-C(z)\ln \lambda] .
\end{eqnarray}

In ref. \cite{MRS} we proposed that by introducing nilpotent
variables a 'superfield' may be defined :
\begin{equation}
\Phi(z,\theta)=C(z)+\theta D(z)
\end{equation}
with the laws of transformation (1) expressed in unison as:
\begin{equation}
\Phi(\lambda z,\theta)=\lambda^{-(\Delta+\theta)} \Phi(z,\theta).
\end{equation}
This equation is equivalent to equation (1), this can easily be
seen by expansion in powers of $\theta$. In this letter we extend
our 'superfield' to a four-component one, exploiting a grassman
variable $\eta$,
\begin{equation}
\Phi(z,\eta)= C(z) + \bar{\alpha}(z) \eta +\bar{\eta} \alpha(z) +
\bar{\eta} \eta D(z).
\end{equation}

Here a fermionic field $\alpha(z)$ with the same conformal
dimension as $C(z)$ has been added to the multiplet, and the
nilpotent variable of ref \cite{MRS} has been interpreted as
$\bar{\eta}\eta$ . Note that both $\alpha$ and $\bar{\alpha}$
live in the holomorphic section of the theory. Now we observe that
$ \Phi (z,\eta )$ has the following transformation law under
scaling

\begin{eqnarray}\label{TRLaw}
\Phi(\lambda z,\eta)&=&\lambda^{-(\Delta+\bar{\eta}\eta)}
\Phi(z,\eta) .
\end{eqnarray}
To find out what this scaling law means, one should expand both
sides of equation(5) in terms of $\eta$ and $\bar{\eta}$. Doing
this and comparing the two sides of equation (5), it is found that
$C(z)$ and $D(z)$ transform as equation (1) and $\alpha$ and
$\bar{\alpha}$ are ordinary fields of dimension $\Delta$. The
appearance of such fields has been proposed by Kausch
\cite{Kausch}, within the $c=-2$ theory.

As discussed in \cite{MRS} using this structure one can derive
most of the properties of LCFTs. Let us first consider the unity
operator. In \cite{MRS} we observed that this structure implies
the existence of logarithmic partners for the unit and the
energy-momentum tensor. The existence of these operators had
previously been noted in the literature \cite{Gur,I1,I2}.

So, in addition to the ordinary unit operator $\Omega$, with the
property $\Omega S = S$ for any field $S$, there exists a
logarithmic partner for $\Omega$, which we denote by $\omega$. We
must have two other fields with zero conformal dimension for the
multiplet to complete:

\begin{equation}
\Phi_0(z,\eta ) = \Omega  + \bar{\xi}(z) \eta +\bar{\eta}
\xi(z)+\bar{\eta} \eta \omega(z).
\end{equation}
with the property that $\langle\Phi_0(z,\eta)\rangle =\bar{\eta}
\eta$. Note that LCFTs have the curious property that $\langle
\Omega\rangle = 0$. The existence of these fields and their OPEs
have been discussed by Kausch \cite{Kausch}. Kausch takes the
ghost action of

\begin{equation}
S=\frac{1}{\pi}\int  d^{2}z
(a\bar{\partial}b+\bar{a}\partial\bar{b})
\end{equation}
with $c=-2$. Instead of the degrees of freedom $a$ and $b$, he
takes two fields $\chi^{\alpha}$ on equal footing, where
$\alpha=1,2$. Taking the Laurent expansion
\begin{equation}
\chi^\alpha =\sum_n \chi^{\alpha}_{n} z^{-n-1}
\end{equation}
he observes that $\chi^{\alpha}_{0}$ plays the role of a ladder
operator for the multiplet defined in equation (6)
\begin{eqnarray}
\chi^{\alpha}_{0}\Omega=0 \hspace{9mm}\\
\chi^{\alpha}_{0}\omega=\xi^{\alpha}\hspace{7mm}\\
\chi^{\alpha}_{0}\xi^{\beta}=d^{\alpha\beta}\Omega,
\end{eqnarray}
 where $d^{\alpha\beta}$ is a totally antisymmetric matrix. Note
 that over the vacuum multiplet we have,
\begin{equation}
\chi^{\alpha}_{0}\chi^{\beta}_{0}=d^{\alpha\beta}L_0.\hspace{2mm}
\end{equation}

An interesting question is whether it is possible to write the
OPEs of the fields in the multiplet in terms of our fields $\Phi_0
(z,\eta)$. The OPE of $\Phi_0$ with itself has to give back
$\Phi_0$ to lowest order:

\begin{equation}
{\Phi}_{0} (z_1,\eta_1) \Phi_0(z_2,\eta_2) \sim
(z_1-z_2)^{\bar{\eta}_1 \eta_2 +\bar{\eta}_2 \eta_1}
\Phi_0(z_1,\eta_3)
\end{equation}
where $\eta_3= \eta_1 + \eta_2$. To see why this OPE has been
proposed, one can look at the behaviour of both sides of
equation(13) under scaling transformation. Under the
transformation $z \rightarrow \lambda z$ the LHS of equation (13)
transforms as $LHS \rightarrow \lambda^{-(\bar{\eta}_1 \eta_1 +
\bar{\eta}_2 \eta_2)} LHS$, (See the transformation law (5)) and
the RHS of equation (13) transforms as $RHS \rightarrow
\lambda^{\bar{\eta}_1 \eta_2 +\bar{\eta}_2 \eta_1- \bar{\eta}_3
\eta_3} RHS$, which are the same. So the OPE proposed here seems
reasonable.

The OPEs of the primary fields involved can then be extracted by
expanding in powers of $\eta$. Taking $z_2=0$ and renaming
$z_1=z$ the OPEs of these fields, aside from trivial OPEs of
$\Omega$, are:

\begin{eqnarray}
\bar{\xi}(z)\xi(0) = 2i(\omega+\Omega\log{z})\hspace{11mm} \\
\xi(z)\omega(0) = - \xi \log{z}\hspace{22mm}\\
\omega(z)\omega(0) = - \log{z} (2\omega+\Omega \log{z})
\end{eqnarray}
These results are consistent with those of \cite{Kausch}. An
obvious generalization of (13) leads to:

\begin{equation}
\Phi_{a}(z_1,\eta_1) \Phi_{b}(z_2,\eta_2) \sim
(z_1-z_2)^{\Delta_{c}- \Delta_{b} - \Delta_{a} + \bar{\eta}_1
\eta_2 + \bar{\eta}_2 \eta_1} C_{abc} \Phi_{c}(z_1,\eta_3)
\end{equation}
for any three primary fields with arbitrary conformal dimensions.
An immediate implication of these equations is that the OPE of two
$C$ fields will be the first components of the multiplets such as
an $\Omega$ or another $C$, thus the expectation values of an
arbitrary string $\langle C(1)C(2)..C(n)\rangle$ will vanish in
any LCFT \cite{FlohrNew}.

A consequence of existence of $\Phi_0$ is the existence of the
energy-momentum multiplet:
\begin{equation}
T(z,\eta)=L_{-2} \Phi_0(z,\eta).
\end{equation}

We thus observe that we have three partners for the
energy-momentum tensor $T(z,\eta)=T_0(z)+\bar{\eta}
\zeta(z)+\bar{\zeta}(z)\eta + \bar{\eta}\eta t(z)$, as discussed
by Gurarie and Ludwig \cite{GL}. In their paper, they have
considered some specific theories and have suggested some OPEs for
different fields of energy-momentum tensor multiplet. However,
getting some insight from the OPEs we have written so far, we
propose
\begin{eqnarray}
T(z,\eta_1)T(0,\eta_2) = \hspace{100mm}\nonumber \\
 z^{ \bar{\eta}_1 \eta_2 + \bar{\eta}_2
\eta_1} \{ \frac{\frac{c(\eta_3)}{2}\Phi_0(\eta_3)}{z^{4}} +
\frac{d(\eta_3)\chi(\eta_3)}{z^3}+
\frac{e(\eta_3)T(\eta_3)}{z^{2}} + \frac{f(\eta_3)
\partial_z{T(\eta_3)}}{z}\}.
\end{eqnarray}

There are some points in this OPE which should be clarified.
First of all, in contrast with the OPE of ordinary
energy-momentum tensor, there exists a $1/z^3$ term. In the
ordinary OPE, this term vanishes because $L_{-1}\Omega=0$. Such a
reasoning can not be extended to the case of other fields of the
energy-momentum multiplet, that is, one can not assume that
$L_{-1}\omega$, $L_{-1}\xi$ and $L_{-1}\bar{\xi}$ all vanish. In
equation (19) we have denoted $L_{-1}\Phi_0(\eta)$ by
$\chi(\eta)=\bar{\eta}\sigma + \bar{\sigma}\eta + \bar{\eta}\eta
J $.

 The other point is that the constants appearing on the RHS
depend only on $\eta_3$. This seems a reasonable assumption. The
constants such as $c(\eta)$ have only two components, that is we
can express them as $c(\eta)=c_1+c_2 \bar{\eta} \eta$, since we
wish to avoid non-scalar constants in our theory. Note that $c_1$
corresponds to the usual central charge in the ordinary theories.
It will be more clear when one looks at the OPE of two $T_0$'s
which is given below. The constant $e(\eta)$ is taken to be
$e(\eta)=2+\bar{\eta} \eta$ for consistency with known conformal
dimension of $T_0(z)$ and with the fact that $t(z)$ is the
logarithmic partner of $T_0(z)$. Also $f(\eta)$ is taken to be
unit, in order to obtain the familiar action of $T_{0}$ on the
members of the multiplet. By now, there is no restrictions on
$d(\eta)$ and we will take it to be $d_1+\bar{\eta}\eta d_2$,
however as shown below, the constant $d_2$ plays no role in the
OPE.

Expanding both sides of equation (19), we find the explicit form
of the OPEs :
\begin{eqnarray}
T_0(z)T_0(0)= \frac{ \frac{c_1}{2} \Omega}{z^{4}}+
\frac{2T_0}{z^{2}}+ \frac{\partial_z T_0}{z} \hspace{88mm} \\
T_0(z)t(0) = \frac{ \frac{c_2}{2} \Omega+
\frac{c_1}{2}\omega}{z^{4}} + \frac{d_1 J}{z^3} + \frac{2t +
T_0}{z^{2}} + \frac{
\partial_z t}{z} \hspace{64mm}\\
t(z)t(0) = -\frac{1}{2} \frac{ \log{z}\left((c_1 \log{z} +
2c_2)\Omega + 2c_1 \omega\right)}{z^{4}} -\frac{2d_1 \log{z} J}{z^3}
- \hspace{40mm} \nonumber \\
\frac{ 2 \log{z}(2t +T_0 \log{z}) + 2 \log{z} T_0}{z^{2}}-\frac{
\partial_z( \log{z}
(2t + T_0 \log{z}))}{z}\hspace{27mm} \\
T_0(z)\zeta(0) = \frac{1}{2} \frac{c_1 \xi}{z^{4}} + \frac{d_1
\sigma}{z^3}+\frac{2
\zeta}{z^{2}}+\frac{\partial_z \zeta}{z}\hspace{80mm}\\
\zeta(z)t(0) = -\frac{1}{2} \frac{c_1 \xi \log{z}}{ z^{4}}
-\frac{d_1 \log{z} \sigma}{z^3}- \frac{2\zeta\log{z}}{z^{2}}-
\frac{\partial_z (\zeta \log{z})}{z}\hspace{43mm}\\
 \bar{\zeta}(z) \zeta(0) = -\frac{1}{2} \frac{ c_1 (\omega +
\Omega \log{z}) + c_2\Omega}{z^{4}}-\frac{d_1 J}{z^3}-\frac{2(t +
T_0 \log{z}) + T_0}{z^{2}}-\frac{\partial_z( t + T_0 \log{z})}{z}.
\end{eqnarray}
We observe that these expressions are not the same as those found
by \cite{GL}. Aside from the presence of the $1/z^3$ terms, the
most important difference is that we have the logarithmic partner
of unity operator in our OPEs as well as unity, itself. The
existence of 'pseudo-unity' when there is a logarithmic partner
for the energy-momentum tensor, is obligatory, in fact one can
easily show that $L_{+2}t(z)$ is the logarithmic partner of
unity. The presence of pseudo-unity causes some problems with the
OPEs derived by \cite{GL}, because now the expectation value of
$\Omega$ is zero. This leads to

\begin{equation}
\langle T_0(z_1)T_0(z_2) \rangle =0
\end{equation}
regardless whether $c_1$ is zero or not. Instead we have
\begin{equation}
\langle T_0(z_1)t(z_2)\rangle = \frac{c_1}{2(z_1-z_2)^4}
\end{equation}
which is just the one written by \cite{GL} with $b$ being $c_1/2$
of our theory. So it is observed that in such theories, there is
no need to set $c_1=0$, however as suggested by \cite{GL}, two
central charges appear as is seen in equation (19). Since this
structure is reminiscent of supersymmetry, most authors have set
$c_1=0$. Setting $c_1$ to zero, we do find a much simpler OPE.
However, two obvious differences with supersymmetry may be
observed. First, $\zeta$ does not transform like the supercurrent
of SCFT. Second, the role of $t(z)$ is not clear. We therefore
believe that connection with supersymmetry, if any, has to appear
at a deeper level. Thus identifying $t$ with the similar object
in \cite{GL} may not be right.

In conclusion it is seen that the idea of grassman variables in
LCFT can be extended to include fermionic fields in the theory,
and this naturally leads to a current algebra involving the
energy momentum tensor, its logarithmic partner and two fermionic
currents. Despite the superficial resemblance to supersymmetry
such as super multiplets etc., there is need of further
clarification if there is any supersymmetry within the theory.
Furthermore the structure of the derived algebra around the
Virasoro algebra remains to be worked out.

\vspace{5mm}

 {\large{\bf Acknowledgements}}

We would like to thank M. Flohr and I. Kogan for helpful comments
on this manuscript. We are also indebted to V. Gurarie for
critical comments on an earlier version of this paper.

\end{document}